\documentclass[a4paper,12pt]{article}

\usepackage{amsmath,amssymb,graphicx}
\usepackage{pst-all}
\usepackage{cite}

\numberwithin{equation}{section}

\newcommand{\ii}{\mathrm{i}}

\newcommand{\dd}{\mathrm{d}}
\newcommand{\pd}{\partial}

\newcommand{\e}{\mathrm{e}}

\newcommand{\Op}{\mathcal{O}}

\newcommand{\Fc}{\check{\Phi}}
\newcommand{\ft}[2]{{\textstyle\frac{#1}{#2}}}
\newcommand{\D}{\mathbf{D}}

\newcommand{\x}{\mathbf{x}}
\newcommand{\y}{\mathbf{y}}
\newcommand{\z}{\mathbf{z}}

\newcommand{\f}[1]{\mathbf{#1}}

\newcommand{\bs}{\backslash}

\begin{document}

\title{Dilatation operator in 3d}%
\author{Corneliu Sochichiu\thanks{On leave from~: Institutul de
    Fizic\u a Aplicat\u a A\c S, str. Academiei, nr. 5,
    Chi\c{s}in\u{a}u, MD2028 Moldova; e-mail:
\texttt{sochichi@sogang.ac.kr}}\\
{\it Center for Quantum Spacetime, }\\
{\it Sogang University,}\\
{\it  Seoul 121-742, Korea}
}%

%

\maketitle
\begin{abstract}
 The perturbative dilatation operator in 3d is constructed at the leading two-loop order.
\end{abstract}
\maketitle
\tableofcontents
\section{Introduction}

Till recently the most information we could have from quantum fields was obtained by Perturbation Theory expansion. Although incomplete in capturing all the properties of the quantum model, the perturbation theory is also very complicated such that our possibilities of exploration just end at few orders. In this situation finding universal patterns or geometrical-like structures in the perturbative expansion is a real challenge, providing the hope of understanding the ``all loop'' behaviour of a quantum field theory model.

On the other hand, from string theory come hints that certain gauge field models may have an equivalent description in terms of \emph{dual} string or gravity theories. This phenomenon is most understood for conformal theories in planar limit and is known as AdS/CFT conjecture \cite{Maldacena:1997re}. According to it, the (super)conformal symmetry corresponds to the (super)isometry of the target space, while the dilatation operator of the conformal theory is mapped to the Hamiltonian of the dual theory. Originally, the AdS/CFT correspondence conjecture was concerning the planar four-dimensional $\mathcal{N}=4$ super Yang--Mills theory and IIB strings on AdS$_5 \times S^5$.

The AdS/CFT correspondence is essentially an Ising-like duality, i.e. a weak coupled super Yang--Mills is mapped to a strong coupled string theory and \emph{viceversa}. However, later it was realized, that there is still a corner in the parameter space of correspondence where both sides can be considered using perturbative tools \cite{Berenstein:2002jq,Tseytlin:2002ny,Frolov:2002av}. This boosted the intensive research in perturbative structure of $\mathcal{N}=4$ super Yang--Mills, which resulted in construction of complete one-loop dilatation operator for this theory \cite{Beisert:2003jj} and discovery of the integrability at the planar level \cite{Minahan:2002ve}. (See \cite{Plefka2006} and \cite{Beisert:2005} for a review of the subject.) In particular, this study lead to formulation of the so called \emph{asymptotic Bethe Ansatz} proposal \cite{Beisert:2005fw}, which summarizes the all-loop structure of the S-matrix up to finite size effects.

Recent proposals regarding multiple M2 brane description
\cite{Bagger:2006sk,Gustavsson:2007,Bagger:2007vi,Bagger:2007jr,Aharony:2008} extended the interest in AdS/CFT correspondence to cases involving three dimensional conformal Chern--Simons theories \cite{Schwarz:2004} as well.
A wide class such theories was found in \cite{Hosomichi:2008jd,Hosomichi:2008jb}. The natural task, which arises is to construct a complete perturbative dilatation operator for this class of theories.

There is also another important motivation for the construction and study of the dilatation operator which, perhaps, is not restricted to conformal theories only, but concerns perturbatively renormalizable theories in general. It consists in the following. Starting from the quantum field theory, one can develop a constructive approach in building `AdS/QFT' correspondence by construction of the dual theory right from the QFT data, by defining the Hamiltonian and Hilbert space of dual theory through the Dilatation operator and correlation functions of the original field theory \cite{Sochichiu:2006uz,Sochichiu:2007am,Sochichiu:2008xd}. For gauge theories this construction leads in the planar limit to string-like models.

Thus, the aim of present work is to provide a frame for construction of the perturbative dilatation operator. It is close in aims and tools to construction of the one-loop four-dimensional operator, proposed in \cite{Sochichiu:2007am}. The difference from the four-dimensional case, though, consists in the fact that the leading contribution to the dilatation operator in the present three-dimensional case comes from \emph{two loops}, since as is well known, there is no log divergence terms at one loop in odd dimensions, while the leading contribution in even dimensions is from one loop. Like in \cite{Sochichiu:2007am}, however, the analysis in the present case does not use any additional symmetry or other assumption rather than the three-dimensional Lorentz invariance and renormalizability. Hence, the results we obtain here can be applied to any renormalizable theory in three dimensions.

The plan of the paper is as follows. In the next section we give the setup for the problem, introduce notations and basic ingredients needed for the computation of the dilatation operator. The `saving' notations are extremely important in this work due to high complexity of the objects. Important element of notations is the use of condensed multi-index, which can be labeled by some irreducible representations of the Lorentz group (or better to say SO(3) group as we work in Euclidean space formulation). In sequent sections we are considering two-, three and four-point contribution to the two-loop dilatation operator, respectively. While analyzing the three vertex level, we use somehow over-condensed notations, which otherwise we avoid to use in the rest of the paper in order to be less obscure. In the Conclusion we summarize the results. For main technical data, computations and useful identities used in the paper the reader is referred to the Appendix.

\section{Setup and notations}

Consider a three-dimensional quantum field theory, described by bosonic fields $\phi(x)$ (condensed notations), fermionic fields $\psi(x)$ and a Chern--Simons gauge field $A_\mu$. The Euclidean Lagrangian of the theory has the following generic form,
\begin{equation}\label{lagrangian}
  L=
  \ft12 (\nabla_\mu\phi)^2+\ii\bar{\psi}\gamma^{\mu}\nabla\psi+
  \ft12 \epsilon^{\mu\nu\lambda}A_\mu\pd_\nu A_\lambda+
  V(\phi,\pd\phi,\psi,\bar{\psi},A).
\end{equation}
This generic form include the conformal Chern-Simons models related to multiple membrane dynamics \cite{Schwarz:2004,Hosomichi:2008jd,Hosomichi:2008jb} as well as other three-dimensional theories of recent interest. In particular, the field $\phi$ can contain also the conventional Yang--Mills component.

In order to perform the perturbative analysis of the model, one has to fix the gauge symmetry. In what concerns the Chern--Simons gauge field, this can be done by addition of the following gauge fixing term \cite{Avdeev:1991za,Avdeev:1992jt},
\begin{equation}
  L_{\rm gf}=(2\xi)^{-1}(\pd_\mu A_\mu)^2,
\end{equation}
and the corresponding Faddeev--Popov ghost field,
\begin{equation}
  L_{\rm gh}=
  \pd_\mu\bar{c}\nabla_\mu c.
\end{equation}
By far, the most convenient choice is the Landau gauge $\xi=0$, in which the propagator for the Chern--Simons field takes the simplest form (see below). For us this simplicity is crucial, since it allows a unified perturbative analysis.

In what concerns the ghost fields, apart from odd Grassmann parity, they follow the same pattern in the Lagrangian, as the bosonic field $\phi$. Therefore, in the following we assume that in addition to ordinary bosons the fields $\phi$ embrace also the gauge ghosts, while the interaction term $V$ is modified correspondingly.

The main subjects of our perturbative analysis are the normal symbols of local composite operators, described in terms of functions of fields and their derivatives at a fixed point, e.g. $x=0$. Since we are interested in the description of operators modulo equations of motion, we can always exclude operators containing the trace parts of derivatives as well as the Chern--Simons fields. In what follows, we assume that the composite operators do not depend on such letters.

Let us turn to the notations. The traceless derivative letters we denote by,
\begin{align}
  \phi_{(\f{n})}&\rightarrow \pd_{(\mu_1}\dots\pd_{\mu_n)}\phi|_{x=0},\\
  \psi_{(\f{n})}&\rightarrow \pd_{(\mu_1}\dots\pd_{\mu_n)}\psi|_{x=0},\qquad
  \bar\psi_{(\f{n})}\rightarrow \pd_{(\mu_1}\dots\pd_{\mu_n)}\bar\psi|_{x=0},
\end{align}
for bosons and fermions respectively. There are no derivative letters appearing for the Chern--Simons field.

In general, we will treat the multi-index $\f n$ as a set: The sum of two multi-index sets $\f n+\f m$, will denote a set of indices containing both those in $\f n$ as well as those in $\f m$. The completion $\f n\bs \f r$ of a multi-index $\f n$ with respect a subset $\f r$ are those indices contained in $\f n$, which are not contained in $\f r$. Also $\f r|\f n$ will denote a partition $\f r\subset \f n$, i.e. a subset of $r$ indices from $\f n$, while $\f r,\f r'|\f n$ correspond to a pair of nonintersecting partitions of $\f n$, with lengths, respectively, $r$ and $r'$.

The Lagrangian \eqref{lagrangian} leads to the following propagators for the letters,
\begin{align}\label{prop:bos}
  \rnode{A}{\phi}_{(\f{n})}\rnode{B}{\phi}_{(\f{m})}
  \ncbar[linewidth=.01,nodesep=2pt,arm=.1,angle=90]{-}{A}{B}
  &=(-1)^{m}\pd_{(\f{n})+(\f{m})}\frac{1}{4\pi x},\\ \label{prop:ferm}
  \rnode{A}{\psi}_{(\f{n})}\rnode{B}{\bar\psi}_{(\f{m})}
  \ncbar[linewidth=.01,nodesep=2pt,arm=.1,angle=90]{-}{A}{B}
  &=(-1)^{m}\gamma^{\f{1}}\pd_{(\f{n})+(\f{m})+\f{1}}\frac{1}{4\pi x},\\ \label{prop:CS}
  \rnode{A}{A}_{\f 1}\rnode{B}{A}_{\f 1'}
  \ncbar[linewidth=.01,nodesep=2pt,arm=.1,angle=90]{-}{A}{B}
  &=\epsilon^{\f 1+\f 1'+\f 1''}\pd_{\f 1''}\frac{1}{4\pi x}=\pd_{\widetilde{\f{11'}}}\frac{1}{4\pi x},
\end{align}
where $x=\sqrt{x^2}$, in \eqref{prop:ferm} there is also a summation over the single index set $\f 1$, while in the Chern--Simons propagator the summation is over $\f 1''$, $\epsilon^{\f 1+\f 1'+\f 1''}$ is the three-dimensional Levi-Civita antisymmetric tensor $\epsilon^{\f 1+\f 1'+\f 1''}\to \epsilon^{\mu\mu'\mu''}$, $\epsilon^{123}=+1$, sometimes the index `dualised' with the antisymmetric tensor we will denote just by dual pair of indices, like $1''\to \widetilde{11'}$ in \eqref{prop:CS}.

In what follows it is convenient to treat all letters as a multi-component field $\Phi=\{\phi_{(\f{n})},\psi_{(\f{n})},\bar\psi_{(\f{n})},A_{\f 1}\}$ with propagator,
\begin{equation}
  \D_x=\rnode{A}{\Phi}_{(\f{n})}(x)\rnode{B}{\Phi}(0)
  \ncbar[linewidth=.01,nodesep=2pt,arm=.1,angle=90]{-}{A}{B}.
\end{equation}

\section{Dilatation operator}

The general form of Dilatation operator in Perturbation Theory is given in \cite{Sochichiu:2007am}. Symbolically, it is represented by the following expression,
\begin{equation}\label{dil-op}
  H=\left[
  \e^{-\int V}*
  \right],
\end{equation}
which denotes the scale dependence of terms arising in OPE of the interaction exponent $\e^{-\int V}$ with a probe composite operator operator. (The notations are essentially explained in \cite{Sochichiu:2007am}, but they can also become clear below during detailed analysis.)

In perturbation theory the dilatation operator can be expanded in powers of interaction potential\footnote{Note the sign difference in the definition of interaction potential with \cite{Sochichiu:2007am}.},
\begin{equation}\label{Delta}
  H=-\int_y [V_{y}*]+
  \frac{1}{2!}\int_{y_1}\int_{y_2} [V_{y_1}* V_{y_2}*]+\dots,
\end{equation}
here and below, the subscript $y$ denotes the integration variable, also we usually will omit the integration measure $\dd^3 y$.

The scale dependence in the square brackets of eqs. \eqref{dil-op} and \eqref{Delta} arise in the following way. The OPE of the terms in the interaction exponent when multiplied to a probe composite operator involves the Wick expansion of the product normal ordered operators which is a source of singular functions. The singular functions need to be regularized and stripped from their singularities. The removal of singularities gives rise to renormalization scale dependence.
In this work we use the, so called, functional form of Wick expansion \cite{Kleinert:1996} and dimensional regularization with minimal subtraction scheme. (See Appendix \ref{app:wick} for further details regarding the functional form of Wick expansion.)

In what follows we analyze the expansion \eqref{Delta} term by term.

\section{One vertex level}
One-vertex level resides in evaluation of divergences of two point functions. At this level we have,
\begin{equation}\label{1-vert-gen}
  H_{\rm one-vertex}=-\int_y[V(y)*]=-\int_y
  \left[
  \e^{\Fc_y\cdot \D_y\cdot \Fc}
  \right]V_y.
\end{equation}
The expansion of the exponential in \eqref{1-vert-gen}, gives respectively the tree level, one-, two-, etc. loop contributions, according to the order. Since $V_y$ is finite order polynomial in $\phi$ of order not higher than six, this series ends up with $k=6$. Also we assume that $V_y$ has at most one one-derivative letter and no higher ones. Before doing the expansion let us rewrite the exponent in \eqref{1-vert-gen} in following form,
\begin{equation}
  \Fc_y\cdot \D_y\cdot \Fc=\sum_{\f n}D_y^{\f n} \check{s}_{\f n},
\end{equation}
where $D^{\f n}_{y}$ is the ${\f n}$-derivative of the fundamental propagator,
\begin{equation}\label{prop-w-tr}
  D^{\f n}_{y}=\pd_{\f n}\frac{1}{4\pi y},
\end{equation}
while the two-point differential $\check{s}^{\f n}_{y0}$ is given by the pairs of letters whose pair correlator is just $D^{\f n}_{y}$. More precisely $\check{s}^{\f n}_{y0}$ is given by,
\begin{multline}\label{s-check}
  \check{s}^{\f n}_{y0}=\\
  \sum_{\substack{\f r,\f l\\ \f r+\f l=\f n}}(-1)^{l}
  \check{\phi}^{(\f r)}_{y} \cdot\check{\phi}^{(\f l)}+
  \sum_{\substack{\f r,\f l, \f 1\\ \f r+\f l+\f 1=\f n}}
  (-1)^{l}\left(\check{\bar{\psi}}^{(\f r)}_{y}\cdot \gamma^{\f 1}\cdot \check{\psi}^{(\f l)}-
  \check{\bar{\psi}}^{(\f r)}\cdot \gamma^{\f 1}\cdot \check{\psi}^{(\f l)}_{y}\right)\\
  +\delta_{\f n,\f 1}\epsilon^{\f 1\f 1' \f 1''} \check{A}_{\f 1'y}\cdot \check{A}_{\f 1''}.
\end{multline}
The last term is non-trivial in the only case when $n=1$.

The derivative letter propagator \eqref{prop-w-tr} may contain both traceless combinations of indices as well as traces. The trace parts in the derivatives will result in contact terms, which we can discard. Therefore, we keep only traceless part in the derivatives.
Taking into account all above, the equation \eqref{1-vert-gen} can be rewritten as,
\begin{equation}
  \int_{y}
  \left[
  \e^{D^{(\f n)}_{y}\check{s}^{(\f n)}_{y0}}\right]V_y=\sum_{k=1}^{6}
  \frac{1}{k!}\int_y
  \left[
  D_{y}^{(\f n_1)}D_{y}^{(\f n_2)}\dots D_{y}^{(\f n_k)}
  \right]\check{s}^{(\f n_1)}_{y0}\check{s}^{(\f n_2)}_{y0}\dots
  \check{s}^{(\f n_k)}_{y0}V_y
\end{equation}

Therefore, the problem is reduced to the computation of the scaling factors,
\begin{equation}\label{1v-notrace}
  \Delta_{(\f{n}_1),(\f{n}_2),\dots,(\f{n}_k)}(x)=\frac{1}{(4\pi)^{k}}
  \left[\prod_{l=1}^{k}\pd_{(\f{n}_l)}\frac{1}{x}\right],
\end{equation}
where in general $k\leq 6$, but since we are limiting ourself to just two loops we consider $k$ only up to three. The scaling factors \eqref{1v-notrace} can be evaluated in many ways. Perhaps, the simplest possibility for the present case is to use the differential renormalization scheme of \cite{Freedman:1991tk} as in \cite{Sochichiu:2007am}. The problem with this scheme is that it is difficult to adapt to higher vertex levels, therefore in the present work we adopt a different although equivalent approach, using the integration technique, based on the method of `uniqueness' \cite{Kazakov:1983ns,Kazakov:1984bw,Kazakov:1984km,Kazakov:1986mu}.

The scaling function \eqref{1v-notrace} for the interesting for us case $k=3$ is evaluated in the Appendix \ref{app:2-point},
\begin{multline}\label{H-1vert}
  H_{\rm one-vertex}=\\
  -\frac{1}{3!(4\pi)^3}\int_y
  \sum_{\f r,s}F_{(\f n),(\f m),(\f k)}^{(\f r),s}\pd_{(\f r)}\pd^{2s}\delta(y)
  \check{s}^{(\f n)}_{y0}\check{s}^{(\f m)}_{y0}
  \check{s}^{(\f k)}_{y0}(V_y)\\
  =-\frac{1}{3!(4\pi)^3}
  \sum_{\substack{\f n,\f m',\f k\\ \f r,s}}\left.(-1)^{r}F_{(\f n),(\f m),(\f k)}^{(\f r),s}\pd_{(\f r)}\pd^{2s}
  \check{s}^{(\f n)}_{y0}\check{s}^{(\f m)}_{y0}
  \check{s}^{(\f k)}_{y0}(V_y)\right|_{y=0},
\end{multline}
where the coefficients $F_{(\f n),(\f m),(\f k)}^{(\f r),s}$ are given by eq. \eqref{F-2point} of the Appendix.

\section{Two vertex level}

Let us turn to the next level. The two-vertex contribution is given by the second term in the operator product expansion,
\begin{equation}\label{h2v}
  H_{\rm 2-vertex}=\frac{1}{2!}\int_{x}\int_y
  \left[
  \e^{\check{\Phi}_{x}\cdot \D_{xy}\cdot\check{\Phi}_{y}+
  \check{\Phi}_{x}\cdot \D_{x}\cdot\check{\Phi}+\check{\Phi}_{y}\cdot \D_{y}\cdot \check{\Phi}}\right]V_{x}V_{y},
\end{equation}
where, as before, the subscript $y$ and $x$ at the derivative letter denotes that the respective letter acts either on $V_{x}$, or on $V_{y}$, which is the interaction potential at the integration point $x$ or $y$, respectively, and the presence of no subscript refers to the point $z=0$.

Using the experience we gained in the previous section, we can conclude that also here the contribution from the trace parts of derivatives acting on a single propagator reduces to a local scale invariant counter-term. Therefore, we can reorganize the r.h.s. of \eqref{h2v} in the following form,
\begin{equation}\label{h2s}
  H_{\rm 2-vertex}=\\
  \frac{1}{2!}\int_{x}\int_y
  \left[
  \e^{D^{(\f n)}_{xy}\check{s}^{(\f n)}_{x,y}+
  D^{(\f n)}_{x0}\check{s}^{(\f n)}_{x 0}+D^{(\f n)}_{y0}\check{s}^{(\f n)}_{y,0}}\right]V_{x}V_{y},
\end{equation}
where $D^{\f n}_{xy}$ and $\check{s}^{\f n}$ are defined in the previous section by equations \eqref{prop-w-tr} and \eqref{s-check}, respectively. Also summation over all allowed $\f n$ is assumed in \eqref{h2s}.
The two-loop contribution is given by the forth term of expansion of the exponent in \eqref{h2v}. The relevant (1pi) terms are,
\begin{multline}\label{h2v2}
  H_{\rm 2-vertex}=
  \int_x\int_y
  \left(
  \ft12[D^{(\f n_1)}_{x}D^{(\f n_2)}_{x}D^{(\f m)}_{y}D^{(\f k)}_{xy}]
  \check{s}^{(\f n_1)}_x \check{s}^{(\f n_2)}_x \check{s}^{(\f m)}_y \check{s}^{(\f k)}_{x,y}\right.\\
  \left.+
  \ft14[D^{(\f n)}_{x}D^{(\f m)}_{y}D^{(\f k_1)}_{xy}D^{(\f k_2)}_{xy}]
  \check{s}^{(\f n)}_x \check{s}^{(\f m)}_y \check{s}^{(\f k_1)}_{x,y}
  \right)V_xV_y,
\end{multline}
and summation over $\f n_{1,2}$, $\f m$, $\f k$ in the first term and $\f n$, $\f m$, $\f k_{1,2}$ in the second one is assumed.

Thus our task reduced to the evaluation of the following scale factors,
\begin{align}\label{delta1}
  \Delta_{(\f n_1)(\f n_2);(\f m);(\f k)}(x,y)&=
  \frac{1}{(4\pi)^4}\left[
  \pd_{(\f n_1)}\frac{1}{x}\pd_{(\f n_2)}\frac{1}{x}
  \pd_{(\f m)}\frac{1}{y}
  \pd^{x}_{(\f k)}\frac{1}{|x-y|}\right],\\ \label{delta2}
  \Delta_{(\f n);(\f m);(\f k_1)(\f k_2)}(x,y)&=
  \frac{1}{(4\pi)^4}\left[
  \pd_{(\f n)}\frac{1}{x}
  \pd_{(\f m)}\frac{1}{y}
  \pd^{x}_{(\f k_1)}\frac{1}{|x-y|}
  \pd^{x}_{(\f k_2)}\frac{1}{|x-y|}\right],
\end{align}
which are linear combinations of products of delta functions of $x$ and $y$  and their derivatives. Moreover, the two scaling factors in \eqref{deltas} satisfy the following relation,
\begin{equation}
  \Delta_{(\f n);(\f m);(\f k_1)(\f k_2)}(x,y)=
  (-1)^{m}\Delta_{(\f k_1)(\f k_2);(\f m);(\f n)}(x-y,-y).
\end{equation}
Therefore in is enough to evaluate only one of the two factors \eqref{delta1} or \eqref{delta2}. This evaluation is done in the Appendix \ref{sec:Three-point-functio-contrib}. The result is given by the following,
\begin{multline}\label{H-2vert}
  H_{\rm 2-vertex}=
  \frac{1}{2(4\pi)^{4}}\times\\
  \left(\sum (-1)^{p+s}
  F^{(\f p)r;(\f s)t}_{(\f n_1)(\f n_2);(\f m);(\f k)}
  (\pd_{(\f p)}\pd^{2r})_x
  (\pd_{(\f s)}\pd^{2t})_y
  \check{s}^{(\f n_1)}_x\check{s}^{(\f n_2)}_x\check{s}^{(\f m)}_y
  \check{s}^{(\f k)}_{xy}\right.\\
  +\left.\ft12\sum (-1)^{m+p+s}
  F^{(\f p)r;(\f s)t}_{(\f k_1)(\f k_2);(\f m);(\f n)}
  (\pd_{(\f p)}\pd^{2r})_x(\pd_{(\f s)}\pd^{2t})_{y}
  \check{s}^{(\f n)}_x
  \check{s}^{(\f n_2)}_x
  \check{s}^{(\f m)}_y
  \check{s}^{(\f k_1)}_{xy}
  \check{s}^{(\f k_2)}_{xy}
  \right)\\
  \times\left.V_xV_y\right|_{x=y=0},
\end{multline}
where the coefficients $F^{(\f p)r;(\f s)t}_{(\f n_1)(\f n_2);(\f m);(\f k)}$ are given by \eqref{F-3point}. The derivatives inside $(~)_x$ or $(~)_y$ are with respect to the subscript variable: $x$ or $y$.

\section{Three vertex level}
Let us move further to the next, the three vertex level. It can be shown that this is the last contribution for the two-loop order. The two-loop contribution at this level is given by scaling factors in the fifth term of the expansion of the expression,
\begin{equation}
  H_{\rm 3-vertex}=-\frac{1}{3!}\int_1\int_2\int_3\left[
  \e^{\check{D}_{1}+\check{D}_{2}+\check{D}_{3}+
  \check{D}_{12}+\check{D}_{13}+\check{D}_{23}}
  \right]
  V_1V_2V_3,
\end{equation}
where we use the extra-shorthand notation $\check{D}_{ij}\equiv D_{x_ix_j}^{(\f n_{ij})}\check{s}^{(\f n_{ij})}_{x_ix_j}=D_{ij}\check{s}_{ij}$.

The relevant terms in the expansion of the exponent are,
\begin{multline}
  \check{D}_{2}\check{D}_{3}\check{D}_{12}\check{D}_{13}\check{D}_{23}+
  \check{D}_{1}\check{D}_{3}\check{D}_{12}\check{D}_{13}\check{D}_{23}+
  \check{D}_{1}\check{D}_{2}\check{D}_{12}\check{D}_{13}\check{D}_{23}\\+
  \check{D}_{1}\check{D}_{2}\check{D}_{3}\check{D}_{13}\check{D}_{23}+
  \check{D}_{1}\check{D}_{2}\check{D}_{2}\check{D}_{12}\check{D}_{23}+
  \check{D}_{1}\check{D}_{2}\check{D}_{3}\check{D}_{12}\check{D}_{23},
\end{multline}
but one can see, that most of these terms are related by permutation of interaction vertices $1,2,3$ and everything reduces to just two distinct contributions with factor three each. Then the three-vertex contribution boils down to
\begin{multline}
  H_{\rm 3-vertex}=\\
  -\frac{1}{2}\int_1\int_2\int_3\left(\left[
  \check{D}_{1}\check{D}_{2}\check{D}_{12}\check{D}_{13}\check{D}_{23}
  \right]+
  \left[
  \check{D}_{1}\check{D}_{2}\check{D}_{3}\check{D}_{12}\check{D}_{23}
  \right]
  \right)V_1V_2V_3.
\end{multline}

Therefore, we are faced to the evaluation of two scale factors,
\begin{subequations}\label{delta3V}
\begin{align}
  \Delta_{(\f n),(\f m);(\f k),(\f l),(\f p)}(x,y,z)&=\\ \nonumber
  &
  \left[
  \pd_{(\f n)}\frac{1}{x}\pd_{(\f m)}\frac{1}{y}
  \pd_{(\f k)}\frac{1}{|x-y|}\pd_{(\f l)}\frac{1}{|x-z|}
  \pd^{y}_{(\f q)}\frac{1}{|y-z|}\right],\\
  \Delta_{(\f n),(\f m),(\f k);(\f l),(\f p)}(x,y,z)&=
  \left[
  \pd_{(\f n)}\frac{1}{x}
  \pd_{(\f m)}\frac{1}{y}
  \pd_{(\f k)}\frac{1}{|x-y|}
  \pd_{(\f l)}\frac{1}{z}
  \pd^{y}_{(\f q)}\frac{1}{|y-z|}\right],
\end{align}
\end{subequations}
of whom, basically, only one is independent due to the the relation,
\begin{equation}\label{duality-3p}
  \Delta_{(\f n),(\f m),(\f k);(\f l),(\f q)}(x,y,z)=
  \Delta_{(\f n),(\f k); (\f m),(\f l),(\f q)}(x,x-y,x-z).
\end{equation}
Therefore, it suffices to evaluate only one scale factor, e.g. $\Delta_{(\f n),(\f m),(\f k);(\f l),(\f q)}(x,y,z)$ and use the duality relation \eqref{duality-3p} to recover another one.

This evaluation is done in the Appendix \ref{sec:Four-point-functio-contrib}, from which we reproduce the result,
\begin{multline}\label{H-3vert}
   H_{\rm 3-vertex}=-\frac{1}{2(4\pi)^5}\biggl\{\\
   (-1)^{p+r+s}F^{(\f p),r;(\f s),t;(\f u),v}_{(\f 1),(\f 2);(\f {12}),(\f {13}),(\f {23})}
   (\pd_{(\f p)}\pd^{2r})_{1}
   (\pd_{(\f s)}\pd^{2t})_{2}
   (\pd_{(\f u)}\pd^{2v})_{3}
   \check{s}_1\check{s}_2\check{s}_{12}\check{s}_{13}\check{s}_{23}\\
   +(-1)^{p}F^{(\f p),r;(\f s),t;(\f u),v}_{(\f 1),(\f {12});(\f 2),(\f 3),(\f {23})}
   (\pd_{(\f p)}\pd^{2r})_{1}
   (\pd_{(\f s)}\pd^{2t})_{2}
   (\pd_{(\f u)}\pd^{2v})_{3}
   \check{s}_{1}\check{s}_{2}\check{s}_{12}\check{s}_{3}\check{s}_{23})
   \biggr\}\\
   \times \left. V_1V_2V_3\right|_{x_1=x_2=x_3=0},
\end{multline}
where the coefficients $F^{(\f p),r;(\f s),t;(\f u),v}_{(\f 1),(\f 2);(\f {12}),(\f {13}),(\f {23})}$ are given by eq. \eqref{F-4point}. The boldface numeric subscripts of $F$'s correspond to multi-indices according to rule: $\f 1\to \f n$, $\f 2\to \f m$, $\f{12}\to \f k$, etc., as we identify $x_1\to x$, $x_2\to y$ and $x_3\to z$.

\section{Discussion}
In this work we constructed the perturbative dilatation operator for a three-dimensional theory, which may include bosonic and fermionic fields as well as Chern--Simons gauge theory. (The last is treated differently from a generic bosonic field due to the specific first order kinetic term.) The result is given by the sum the right hand sides of eqs. \eqref{H-1vert}, \eqref{H-2vert} and \eqref{H-3vert}, which represent the contribution of, respectively, 2-, 3-, and 4-point functions at the two-loop level. As there is no one-loop contribution to the dilatation operator, this is the leading order. It is worth noting, that no additional assumptions including conformal or other symmetry was made. Therefore the results can be applied to any renormalizable model in three dimensions. In the case of existence of additional symmetries and, in particular, (super)conformal symmetry the final form of dilatation operator is expected to simplify considerably.

The dilatation operator has the following universal structure: It is a linear combination of differential operators which depend only on the field content of the theory with coefficients, which we call scaling coefficients, which carry the information of analytical properties of three dimensional space and are independent of the theory. This means that such coefficients need to be evaluated only once for all three dimensional models, what we have actually done in this work, while all model dependence is encoded in some differential operator quantities.

The subjects of the dilatation operator are local composite operators, which should also be gauge invariant. The equations of motion can be used to set such operators into a \emph{canonical form}, which includes only traceless derivative letters and no Chern--Simons gauge fields. Our formulation assumes, that such canonical local composite operators are used as input for the dilatation operator. However, as the result of the action of dilatation operator contains the traced derivatives, the output, generically, is not given in the canonical form, i.e. it may contain the traces of derivatives as well as Chern--Simons gauge fields. Here we did not find any elegant way to reduce the result to the canonical form for the general case, so we leave this problem to future research.

Another topic left for future research is the application of the obtained results to the ABJM or more general models of \cite{Aharony:2008,Hosomichi:2008jd,Hosomichi:2008jb}. In particular, it would be interesting to reproduce in the planar limit the integrable Hamiltonians of \cite{Minahan:2008hf,Bak:2008cp,Bak:2008vd}.

\subsection*{Acknowledgements}
I benefited from useful discussions with Dongsu Bak, Chanyong Park and Ren\'{e} Meyer. I am grateful to my colleagues by CQUeST and in particular to Bum-Hoon Lee, Jeong-Hyuck Park and Hyeong-Chan Kim for friendly, creative atmosphere and interest in my research.

This work was supported by Center for Quantum Spacetime (CQUeST) of Sogang University with grant number R11-2005-021.
\appendix
\section{Differential form of Wick expansion}\label{app:wick}
Here we give the derivation of the differential form of Wick expansion. Since we are interested in the formula for the particular case concerning the product of two and more normal ordered \emph{local} operators, we derive it strictly for this case. A general derivation can be found in  \cite{Kleinert:1996}.

Consider first the case of two operators,
\begin{equation}
  :\Op_1::\Op_2:,
\end{equation}
where for saving notations 1,2 denote not only distinct operators but also distinct space-time points. Thus, each of these operators can be represented as a combination of functional derivatives with respect to $J$ at $J=0$ of the generating functional,
\begin{equation}
  :\e^{\ii J\Phi}:,
\end{equation}
where $\Phi$ can represent arbitrary letter, including derivative ones.
Thus, it suffices to evaluate the Wick expansion for,
\begin{equation}
  :\e^{\ii J_1\Phi_1}::\e^{\ii J_2\Phi_2}:.
\end{equation}
To do this, let us split the field $\Phi$ in positive and negative modes $\Phi^{\pm}$, such that,
\begin{equation}
  \rnode{A}{\Phi}\rnode{B}{\Phi}_{xy}
  \ncbar[linewidth=.01,nodesep=2pt,arm=.1,angle=90]{-}{A}{B}
  =[\Phi^{-}_x,\Phi^{+}_{y}].
\end{equation}

Using this expansion and the Baker--Campbell--Hausdorff formula in the form,
\begin{equation}
  \e^{A}\e^{B}=\e^{[A,B]}\e^{B}\e^{A},
\end{equation}
for two operators $A$ and $B$ with scalar commutator, we have
\begin{multline}\label{wick2p}
  \e^{\ii J_1\Phi^{+}_1}\e^{\ii J_1\Phi^{-}_1}\e^{\ii J_2\Phi^{+}_2}\e^{\ii J_2\Phi^{-}_2}=
  \e^{\mp J_1[\Phi^{-}_1,\Phi^{+}_2]J_2}\e^{\ii J_1\Phi^{+}_1}\e^{\ii J_2\Phi^{+}_2}\e^{\ii J_1\Phi^{-}_1}\e^{\ii J_2\Phi^{-}_2}\\
  \equiv
  \e^{\check{\Phi}_1 \cdot\D_{12}\cdot\check{\Phi}_2}
  \left(
  :\Op_1\Op_2:
  \right),
\end{multline}
where we used the correspondence in the replacement of $J$ with $-\ii \check{\Phi}$.

To get the right sign for fermionic contribution the fermionic derivatives should be properly defined. We use the left derivative for the fermion and right one for the anti-fermion. Note also, that the propagator in \eqref{wick2p} is the correlator of derivative letters, therefore it is given by a number of derivatives acting on the first argument of the propagator for the fundamental letters and another number acting on the second one. Since the fundamental propagator depends only on the difference of the arguments, this can be expressed up to a sign factor as derivatives with respect to the first factor only. Any trace part of the derivatives will produce a contact term. The contact terms lead to bubbling  of some loops, therefore they can be discarded. As a result we can discard any trace part in the propagator $\D_{12}$, while leaving only the trace-less part of derivatives.

The exponents contains the sum of pairs of derivative letters. Each term has the form $\check{\Phi}_x\cdot\D_{xy}\cdot\check{\Phi}_y$, and can be recast into the form,
\begin{equation}
  \check{\Phi}_x\cdot\D_{xy}\cdot\check{\Phi}_y=
  \sum_{(\f n)}D^{(\f n)}_{xy}\check{s}^{(\f n)}_{xy}\equiv
  \check{D}_{xy},
\end{equation}
where $D^{(\f n)}_{xy}$ are given by \eqref{prop-w-tr} and $\check{s}^{(\f n)}_{xy}$ by \eqref{s-check}.

The Wick expansions of the product of three and more factors are given by the following expressions,
\begin{align}
  :\Op_1::\Op_2::\Op_3:&=
  \e^{\check{D}_{12}+\check{D}_{13}+\check{D}_{23}}
  :\Op_1\Op_2\Op_3:,\\
  :\Op_1::\Op_2::\Op_3::\Op_4:&=
  \e^{\check{D}_{12}+\check{D}_{13}+\check{D}_{23}+\check{D}_{14}
  +\check{D}_{24}+\check{D}_{34}}
  :\Op_1\Op_2\Op_3\Op_4:,\\
  :\Op_1::\Op_2:\dots :\Op_k:&=
  \e^{\left(
  \sum_{l<m}\check{D}_{lm}
  \right)}
  :\Op_1\Op_2\dots\Op_k:.
\end{align}

\section{Useful identities}\label{app:useful}
Here we give some useful identities used to compute the scaling factors. Most of these identities can be found in \cite{Chetyrkin:1980pr,Kazakov:1983ns,Kazakov:1984bw,Kazakov:1984km,Kazakov:1986mu}.

An identity we extensively use in this work is related to the traceless derivative of $\pd_{(\f n)}1/x^{\nu}$. It can be expressed as follows,
\begin{equation}
  \pd_{(\f n)}\frac{1}{x^{\nu}}=
  (-1)^{n}\frac{\Gamma\left(\ft{\nu}{2}+n\right)}{\Gamma\left(\ft{\nu}{2}\right)}
  \frac{\x^{(\f n)}}{x^{\nu+2 n}}=(-1)^{n}
  \frac{(\nu+2n-2)!!}{(\nu-2)!!}
  \frac{\x^{(\f n)}}{x^{\nu+2n}}.
\end{equation}
This identity can be obtained by representing $x^{-\nu}$ as,
\begin{equation}
  \frac{1}{x^{\nu}}=\frac{1}{\Gamma\left(\ft{\nu}{2}\right)}\int_0^{+\infty}\dd t\, t^{\frac{\nu}{2}-1}\e^{-x^2 t},
\end{equation}
and applying the derivative. A related identity expresses  $1/x^{\nu}$ as Fourier transform,
\begin{equation}\label{Fourier}
  \int\frac{\dd^D p}{p^{\nu}}\,\e^{\ii px}=\frac{\Gamma\left(\ft{D-\nu}{2}\right)}{\Gamma\left(\ft{\nu}{2}\right)}
  \frac{2^{D-\nu}\pi^{D/2}}{x^{D-\nu}},
\end{equation}
where $D=3-\epsilon$ is the (analytic) space dimension.

Another important identities are related to the decomposition of the monomial $x^{\f n}$ into traceless parts and squares,
\begin{multline}\label{id-beta}
  \x^{\f n}=
  \sum_{\substack{\f r,\f r'| \f n\\ r=r'}}
  \frac{\Gamma(n-2r+D/2)2^{-2r}}{\Gamma(n-2r+1)\Gamma(n-r+D/2)}
  g^{\f r,\f r'}\x^{(\f n\bs\f r\bs\f r')}x^{2r}\\
  \equiv
  \sum_{\f r''| \f n}
  \beta^{\f n}_{(\f n\bs\f r'')}\x^{(\f n\bs\f r'')}x^{2r''},
\end{multline}
where $g^{\f r,\f r'}$ represents the product of metric tensors with indices given by corresponding elements from $\f r$ and $\f r'$ respectively. The ``inverse'' transformation is given by,
\begin{equation}\label{id-beta-inv}
  \x^{(\f n)}=
  \sum_{\substack{\f r,\f r'| \f n\\ r=r'}}
  \frac{\Gamma(n-r+D/2-1)2^{-2r}}{\Gamma(n-2r+1)\Gamma(n+D/2-1)}g^{\f r,\f r'}\x^{\f n\bs\f r\bs\f r'}x^{2r}.
\end{equation}
In ``conventional'' form the identities \eqref{id-beta} and \eqref{id-beta-inv} are given, respectively, by,
\begin{subequations}
\begin{align}
  \nonumber
  x^{\mu_1}\dots x^{\mu_n}&=
  \hat{S}\sum_{r=0}^{[n/2]}
  \frac{\Gamma(n-2r+D/2)2^{-2r}n!}{\Gamma(n-2r+1)\Gamma(n-r+D/2)}\\
  &\qquad\times
  g^{\mu_1\mu_2}\dots g^{\mu_{2r-1\mu_{2p}}}x^{(\mu_{2r+1}\dots \mu_n)}x^{2r},\\ \nonumber
  x^{(\mu_1\dots\mu_n)}&=
  \hat{S}\sum_{p\geq 0}
  \frac{\Gamma(n-r+D/2-1)2^{-2r}n!}{\Gamma(n-2r+1)\Gamma(n+D/2-1)r!}\\
  &\qquad\times
  g^{\mu_1\mu_2}\dots g^{\mu_{2r-1}\mu_{2r}}x^{\mu_{2r+1}}\dots x^{\mu_n}x^{2r},
\end{align}
\end{subequations}
where $\hat{S}$ is the index symmetrization operator.

\section{Computation of the two point contribution}\label{app:2-point}

A generic two-point contribution looks like,
\begin{equation}\label{2-point}
  \Delta_{(\f n),(\f m),(\f k)}(x)=\frac{1}{(4\pi)^{3}}
  \left[
  \pd_{(\f n)}\frac{1}{x}
  \pd_{(\f m)}\frac{1}{x}
  \pd_{(\f k)}\frac{1}{x}
  \right]
\end{equation}

Since the theory is renormalizable, the counter-terms and, therefore, the scaling factors have a local nature. Therefore, the scaling function \eqref{2-point} has to be a combination of $\delta$-function and its derivatives. The generic form of this is given by,
\begin{equation}
  \Delta_{(\f n),(\f m),(\f k)}(x)=\frac{1}{(4\pi)^{3}}
  \sum_{\f r,s}F_{(\f n),(\f m),(\f k)}^{(\f r),s}\pd_{(\f r)}\pd^{2s}\delta(x),
\end{equation}
where the coefficients $F_{(\f n),(\f m),(\f k)}^{(\f r),s}$ are defined by,
\begin{equation}\label{f-rs}
  F_{(\f n),(\f m),(\f k)}^{(\f r),s}=f^{(\f r),s}_{nmk}
  \left[\int_{x}\frac{\x^{(\f n)+(\f m)+(\f k)+(\f r)}}{x^{3+2(n+m+k-s)}}\right],
\end{equation}
where
\begin{equation}
  f^{(\f r),s}_{nmk}=
  (-1)^{n+m+k}(2n-1)!!(2m-1)!!(2k-1)!!
  \alpha^{(\f r),s}
\end{equation}
and the factors $\alpha^{(\f n),r}$ are the trace-reduced coefficients of Taylor expansion,
\begin{equation}
  V_{x}=\sum_{(\f n),r}\alpha^{(\f n),r}\x^{(\f n)}x^{2r}\pd_{(\f n)}\pd^{2r}V_{0}.
\end{equation}

The integral in the square brackets of \eqref{f-rs} diverge in both UV ($x\to 0$) and IR ($x\to\infty$). The IR regularization can be introduced by a mass-like decay-off of the integrand. In this case the UV divergence is taken care of by the extension of the dimension of the integral to analytical value of $D$,
\begin{equation}\label{f-rs-int}
  \left[\int_{x}\frac{\x^{(\f n)+(\f m)+(\f k)+(\f r)}}{x^{3+2(n+m+k-s)}}\right]_{reg}\equiv
  \int\dd^{D}x\,\frac{\x^{(\f n)+(\f m)+(\f k)+(\f r)}}{x^{3+2(n+m+k-s)}}\e^{-\mu x}.
\end{equation}
Then, the desired scale dependence is given by the scale ($\log\mu$) dependence of the UV finite part.

The integral in the r.h.s of \eqref{f-rs-int} can be evaluated as follows. Let us separate the integration into the radial integration $\dd x$ and the angular part $\dd \hat{x}$. Then,
the integral becomes a product of radial and angular integrals,
\begin{multline}\label{2p-scal}
  \int\dd^{D}x\,\frac{\x^{(\f n)+(\f m)+(\f k)+(\f r)}}{x^{3+2(n+m+k-s)}}\e^{-\mu x}\\
  =\beta^{(\f n)+(\f m)+(\f k)+(\f r)}\int_0^{\infty}\dd x\,
  x^{D-(4+n+m+k)+r+2s}\e^{-\mu x}=\\
  \mu^{3-D+(n+m+k)-r-2s}\beta^{(\f n)+(\f m)+(\f k)+(\f r)}\Gamma(D-(3+n+m+k)+r+2s),
\end{multline}
where,
\begin{equation}
  \beta^{\f n}=
  \int_{\hat{\x}^2 =1}\dd\hat{\x}\,\hat{\x}^{\f n}=
  \frac{2^{-(n+1)}n!\pi^{D/2}}{\Gamma\left(\ft{D+n}{2}\right)}
  g^{\f n},
\end{equation}
with $g^{\f n}=0$ for odd $n$ while for even $n$ it is the symmetrized product of metric tensors,
\begin{equation}
  g^{\f n}\mapsto \frac{1}{|S_{n}|}\sum_{p\in S_{n}}g^{\mu_{p(1)}\mu_{p(2)}}\dots g^{\mu_{p(n-1)}\mu_{p(n)}}.
\end{equation}

The relevant contribution comes at $(n+m+k)-r-2s=0$, and it is given by,
\begin{equation}
  \frac{2^{-(n+m+k+1)}(n+m+k)!\pi^{D/2}}{\Gamma\left(\ft{D+n+m+k}{2}\right)}
  g^{(\f n)+(\f m)+(\f k)}\mu^{-\epsilon}\Gamma(\epsilon).
\end{equation}

Then, the scaling factor is given by the coefficient in front of $\log\mu$ term for small $\mu$ expansion, i.e.,
\begin{equation}\label{F-2point}
  F_{(\f n),(\f m),(\f k)}^{(\f r),s}=-\delta_{n+m+k-r-2s,0}f^{(\f r),s}_{nmk}
  g^{(\f n)+(\f m)+(\f k)}
  \frac{2^{-\ft{n+m+k}{2}}\pi}{(n+m+k+1)!!}
\end{equation}
\section{Three point function contribution}\label{sec:Three-point-functio-contrib}

Consider the contribution,
\begin{equation}\label{Dn1n2mk}
  \Delta_{(\f n_1),(\f n_2);(\f m);(\f k)}(x,y)=\frac{1}{(4\pi)^4}
  \left[
  \pd_{(\f n_1)}\frac{1}{x}\pd_{(\f n_2)}\frac{1}{x}
  \pd_{(\f m)}\frac{1}{y}
  \pd_{(\f k)}\frac{1}{|x-y|}
  \right].
\end{equation}

From general considerations, the structure of the scaling function \eqref{Dn1n2mk} is given by,
\begin{equation}
  \Delta_{(\f n_1),(\f n_2);(\f m);(\f k)}(x,y)=\frac{1}{(4\pi)^4}
  \sum_{\substack{\f p,r\\ \f s,t}} F^{(\f p),r;(\f s),t}_{(\f n_1),(\f n_2);(\f m);(\f k)}
  \pd_{(\f p)}\pd^{2r}\delta(x)
  \pd_{(\f s)}\pd^{2t}\delta(y),
\end{equation}
where the coefficients $F^{(\f p),r;(\f s),t}_{(\f n_1),(\f n_2);(\f m);(\f k)}$ are given by the scale flow of the following integral,
\begin{multline}\label{Fprst}
  F^{(\f p),r;(\f s),t}_{(\f n_1),(\f n_2);(\f m);(\f k)}=\\
  \alpha^{(\f p),r}\alpha^{(\f s),t}\int_x\int_y
  \left[
  \pd_{(\f n_1)}\frac{1}{x}\pd_{(\f n_2)}\frac{1}{x}
  \pd_{(\f m)}\frac{1}{y}
  \pd_{(\f k)}\frac{1}{|x-y|}
  \x^{(\f p)}x^{2r}\y^{(\f s)}y^{2t}
  \right].
\end{multline}

The coefficients \eqref{Fprst} can be evaluated as follows,
\begin{equation}\label{Fprst-ev1}
  F^{(\f p),r;(\f s),t}_{(\f n_1),(\f n_2);(\f m);(\f k)}=
  f_{n_1n_2mk}\int_x\int_y
  \left[
  \frac{\x^{(\f n_1)+(\f n_2)+(\f p)}}{x^{2+2(n_1+n_2-r)}}
  \frac{\y^{(\f m)+(\f s)}}{y^{1+2(m-t)}}
  \frac{(\x-\y)^{(\f k)}}{|x-y|^{1+2k}}
  \right].
\end{equation}
where,
\begin{multline}
  f_{n_1n_2mk}=\\
  (-1)^{n_1+n_2+m+k}(2n_1-1)!!(2n_2-1)!!(2m-1)!!(2k-1)!!\alpha^{(\f p),r}\alpha^{(\f s),t}
\end{multline}

A powerful technique to compute Feynman diagrams, dubbed `method of uniqueness' was developed in
\cite{Kazakov:1983ns,Kazakov:1984bw,Kazakov:1984km,Kazakov:1986mu}. The idea of the method consists in reduction of complicated Feynman diagrams to simpler ones using their analytical properties and duality relations among them. One, particularly useful trick is the possibility to express the contribution of a chain of propagators as a single propagator contribution, using the identity,
\begin{equation}\label{uniq1}
  \int_y\frac{(\x-\y)^{\f m}}{|x-y|^{\beta}}\frac{\y^{\f n}}{y^{\alpha}}=v_{nm}(\alpha,\beta)\frac{\x^{\f n+\f m}}{x^{\alpha+\beta-D}},
\end{equation}
where the coefficient $v_{nm}(\alpha,\beta)$ is given by
\begin{equation}
  v_{nm}(\alpha,\beta)=
  \pi^{D/2}2^{-2(n+m)}
  \frac{\Gamma(\ft{D-\alpha}{2}+n)\Gamma(\ft{D-\beta}{2}+m)\Gamma(\ft{\alpha+\beta-D}{2})}
  {\Gamma(\ft{\alpha}{2})\Gamma(\ft{\beta}{2})\Gamma(D-\ft{\alpha+\beta}{2}+m+n)}.
\end{equation}
Let us note, that for $\alpha+\beta\geq 2$ the above coefficients are \emph{regular}, while for even $\alpha+\beta\geq 6$ they vanish in the limit $D\to 3$, due to the Gamma-function pole in the denominator.

The relation \eqref{uniq1} can be obtained by using the Fourier transform \eqref{Fourier} and observing that the convolution in real space becomes an ordinary product in momentum space.

Taking into account the identity \eqref{uniq1}, the coefficient function \eqref{Fprst-ev1} can be equivalently rewritten in the form,
\begin{multline}\label{Fprst-uniq}
  F^{(\f p),r;(\f s),t}_{(\f n_1),(\f n_2);(\f m);(\f k)}=f_{n_1n_2mk}\\
  \times\left[v_{k,m+s}(1+2(m-t),1+2k)\int_x
  \frac{\x^{(\f n_1)+(\f n_2)+(\f p)+(\f m)+(\f s)+(\f k)}}{x^{4+2(n_1+n_2-r+m-t+k)-D}}
  \right].
\end{multline}
The regularized version of the divergent integral in the r.h.s of \eqref{Fprst-uniq} was already evaluated in the section \ref{app:2-point} of the Appendix. It is given by,
\begin{multline}
  \left[
  \int_x
  \frac{\x^{(\f n_1)+(\f n_2)+(\f p)+(\f m)+(\f s)+(\f k)}}{x^{4+2(n_1+n_2-r+m-t+k)-2D}}
  \right]_{reg}=\\
  \mu^{3-D+(n_1+n_2+m+k)-(p+s+2r+2t)}\beta^{(\f n_1)+(\f n_2)+(\f m)+(\f k)+(\f p)+(\f s)}\times\\
  \Gamma(D-(3+n_1+n_2+m+k)+p+s+2r+2t).
\end{multline}
The relevant contribution to the scaling coefficients comes at $(n_1+n_2+m+k)-(p+s+2r+2t)=0$. Therefore the scaling factors \eqref{Fprst-ev1} are given by
\begin{multline}\label{F-3point}
  F^{(\f p),r;(\f s),t}_{(\f n_1),(\f n_2);(\f m);(\f k)}=
  -\delta_{(n_1+n_2+m+k),(p+s+2r+2t)}f_{n_1n_2mk}\\ \times
  \left.v_{k,m+s}(1+2(m-t),1+2k)\beta^{(\f n_1)+(\f n_2)+(\f m)+(\f k)+(\f p)+(\f s)}\right|_{D=3}.
\end{multline}

\section{Four point function contribution}\label{sec:Four-point-functio-contrib}

Consider the four-point function,
\begin{multline}
  \Delta_{(\f n),(\f m);(\f k),(\f l),(\f p)}(x,y,z)=\\
  \frac{1}{(4\pi)^{5}}\left[
  \pd_{(\f n)}\frac{1}{x}\pd_{(\f m)}\frac{1}{y}
  \pd_{(\f k)}\frac{1}{z}\pd_{(\f l)}\frac{1}{|x-z|}
  \pd^{y}_{(\f q)}\frac{1}{|y-z|}\right].
\end{multline}

Similarly to the previous case, we can consider the general structure of the dilatation operator,
\begin{multline}\label{V3structure}
  \Delta_{(\f n),(\f m);(\f k),(\f l),(\f p)}(x,y,z)=\\
  \frac{1}{(4\pi)^{5}}
  \sum_{\substack{\f p,r\\ \f s, t \\ \f u, v}}
  F^{(\f p),r;(\f s),t;(\f u),v}_{(\f n),(\f m);(\f k),(\f l),(\f q)}
  \pd_{(\f p)}\pd^{2r}\delta(x)
  \pd_{(\f s)}\pd^{2t}\delta(y)
  \pd_{(\f u)}\pd^{2v}\delta(z),
\end{multline}
and let us evaluate the coefficients $F$.

The scaling factors $F^{(\f p),r;(\f s),t;(\f u),v}_{(\f n),(\f m);(\f k),(\f l),(\f q)}$ are given by,
\begin{multline}\label{Fprstuv}
  F^{(\f p),r;(\f s),t;(\f u),v}_{(\f n),(\f m);(\f k),(\f l),(\f q)}=
  f_{nmklq}^{(\f p),r;(\f s),t;(\f u),v}\\ \times
  \int_x\int_y\int_z
  \left[
  \frac{\x^{(\f n)+(\f p)}}{x^{1+2(n-r)}}
  \frac{\y^{(\f m)+(\f s)}}{y^{1+2(m-t)}}
  \frac{\z^{(\f k)+(\f u)}}{z^{1+2(k-v)}}
  \frac{(\x-\z)^{(\f l)}}{|x-z|^{1+2l}}
  \frac{(\y-\z)^{(\f q)}}{|y-z|^{1+2q}}
  \right],
\end{multline}
where
\begin{multline}
  f_{nmklq}^{(\f p),r;(\f s),t;(\f u),v}=
  (-1)^{n+m+k+l+q}\times\\
  (2n-1)!!(2m-1)!!(2k-1)!!(2l-1)!!(2q-1)!!
  \alpha^{(\f p),r}\alpha^{(\f s),s}\alpha^{(\f u),v}
\end{multline}

The scaling coefficient \eqref{Fprstuv} too, can be reduced to the two point function expression, which was already evaluated in the section \ref{app:2-point} of the present Appendix. This is done by integration over $x$ and $y$ using the identity \eqref{uniq1}. As a result we have,
\begin{multline}\label{4-point}
  \left[
  \frac{\x^{(\f n)+(\f p)}}{x^{1+2(n-r)}}
  \frac{\y^{(\f m)+(\f s)}}{y^{1+2(m-t)}}
  \frac{\z^{(\f k)+(\f u)}}{z^{1+2(k-v)}}
  \frac{(\x-\z)^{(\f l)}}{|x-z|^{1+2l}}
  \frac{(\y-\z)^{(\f q)}}{|y-z|^{1+2q}}
  \right]=\\
  (-1)^{q+l}v_{q,m+s}(1+2q,1+2(m-t))v_{l,n+p}(1+2l,1+2(n-r)) \\
  \times
  \left[\int_{z}\frac{\z^{(\f n)+(\f m)+(\f k)+(\f l)+(\f q)+(\f s)+(\f p)+(\f u)}}{z^{5+2(n+m+k+l+q-r-t-v)-2D}}\right]
\end{multline}
The regularized value of the integral in the last line of \eqref{4-point} was found in \eqref{2p-scal}. It appears to be,
\begin{multline}
  \left[\int_z\frac{\z^{(\f n)+(\f m)+(\f k)+(\f l)+(\f q)+(\f s)+(\f p)+(\f u)}}{z^{5+2(n+m+k+l+q-r-t-v)-2D}}\right]=
  \beta^{(\f n)+(\f m)+(\f k)+(\f l)+(\f q)+(\f s)+(\f p)+(\f u)}\times \\
  \mu^{-3D+4+n+m+k+l+q-2(r+t+v))}\Gamma(3D-4-(n+m+k+l+q)+2(r+t+v)).
\end{multline}
The contribution to the dilatation operator comes when the power of $\mu$ vanishes in the limit $D\to 3$. This happens when $n+m+k+l+q=2(r+t+v)+5$. Therefore, the scaling coefficient becomes,
\begin{multline}\label{F-4point}
   F^{(\f p),r;(\f s),t;(\f u),v}_{(\f n),(\f m);(\f k),(\f l),(\f q)}=\delta_{n+m+k+l+q,2(r+t+v)+5}
  f_{nmklq}^{(\f p),r;(\f s),t;(\f u),v}(-1)^{q+l}\\
  \times v_{q,m+s}(1+2q,1+2(m-t))v_{l,n+p}(1+2l,1+2(n-r))\\
  \times\left.
  \beta^{(\f n)+(\f m)+(\f k)+(\f l)+(\f q)+(\f s)+(\f p)+(\f u)}
  \right|_{D=3}.
\end{multline}

\bibliographystyle{hunsrt}
\bibliography{adscft}
\end{document}